\documentclass[a4paper]{jpconf}
\usepackage{graphicx}
\usepackage{graphicx,amssymb,amsmath}

\begin{document}
	
\title{Primordial black holes and the Sunyaev-Zel'dovich effect}

\author{Justine Tarrant and Geoff Beck}

\address{University of the Witwatersrand, 1 Jan Smuts Avenue, Braamfontein, South Africa, 2000}

\ead{justine.tarrant@wits.ac.za, geoffrey.beck@wits.ac.za}

\begin{abstract}
Primordial black holes are a popular candidate for dark matter. In the mass regime where their
conjectured Hawking evaporation is significant, they have been subject to many constraints via X-rays,
gamma-rays, and even radio emission. Previously the Sunyaev-Zel'dovich effect has been
considered to place further limits on the $M>10 M_\odot$ primordial black hole abundance via the effects of their
accretion of ambient gas. In this work we will present a novel and robust means of placing abundance limits on low-mass black holes, using the
Sunyaev-Zel'dovich effect induced by electrons produced via their Hawking radiation within galaxy clusters.
\end{abstract}

\section{Introduction}

Recently, the study of primordial black holes (PBHs) has gained significant attention due to the possibility that they could constitute dark matter (DM)~\cite{Zeldovich:Novikov,Hawking}. Aside from its gravitational interaction, very little is known about DM. PBHs interact almost entirely via gravitation, as does DM. This strengthens the link between DM and PBHs and the possibility of a PBH-DM fraction that is non-zero. PBHs are unique in that they are not formed from stellar collapse. They are created in the early universe within the first second after the big bang. Furthermore, the mass of the PBHs depends on the time at which they were created. Those that formed at the Planck time ($ t\sim 10^{-43}$s) have masses around $10^{-5}$g, whilst a mass of $10^{38}$g occurs at $t\sim 1$s. This broad mass range might indicate that extended mass functions are more realistic~\cite{Boudaud:Cirelli}. The mass range considered here is subject to astrophysical and cosmological constraints. For example, large masses ($m\geq 10^{38}$g) are subject to dynamical constraints and strong lensing~\cite{Brandt:2016,Koushiappas:2017,Gaggero:2017}, whilst masses below $m=4\times 10^{14}$g are excluded because they would have already evaporated due to Hawking radiation. Observations using the Voyager $1$ spacecraft~\cite{Boudaud:Cirelli}, lensing~\cite{Chan:Lee,Montero_Camacho} and gamma-rays~\cite{CarrKohri:2016,Carr:2020} have been used to further constrain PBH-DM abundance~\cite{Chan:Lee}. Therefore, we consider the range of masses: $m \in [4\times 10^{14},10^{17}]$g. The upper bound in this mass range results from the Hawking flux being very small for $m>10^{17}$g.

Black holes are conjectured to undergo Hawking radiation, as do PBHs, and when these black holes are large they emit low-energy charged particles like electrons. At sub-GeV energies, these electrons are influenced by the Sun~\cite{Boudaud:Cirelli}. This meant that probing the low-energy scales was not feasible. As already mentioned, the low-energy electrons correspond to large masses meaning that it is difficult to determine PBH abundance for larger masses~\cite{Boudaud:Cirelli}. Fortunately, the Voyager $1$ spacecraft has left the heliopause threshold, making it possible to detect low-energy electrons, which are no longer being heavily influenced by the Sun~\cite{Voyager:2013,Voyager:2016}. Boudad \& Cirelli 2019 \cite{Boudaud:Cirelli}, in their work, used the Voyager $1$ data to place constraints on the PBH abundance. 

On the other hand, when PBHs are small enough they have a high temperature~\cite{Hawking:PBH}. This means that they can emit a population of fast, energetic electrons. These electrons from PBH evaporation are likely to emit synchrotron radiation in the presence of a magnetic field~\cite{Chan:Lee}. Chan \& Lee 2020~\cite{Chan:Lee} focus on radio data (and synchrotron radiation) from the inner Galactic Centre. Finally, one may also place constraints on the PBH abundance using the SZ effect to study the effects of the accretion of ambient gas~\cite{Abe:2019}. The SZ effect considers cosmic microwave background photons that get up-scattered by energetic electrons in the intervening space. This has the virtue of being far less uncertain than the method of Chan \& Lee, which depends very strongly upon the assumptions used to derive the magnetic field profile within the Milky-Way galactic centre (for which only a few data points and weak theoretical constraints exist~\cite{Bringmann:2014}). In addition to this, a single radio data point is all that is used for synchrotron constraints~\cite{Chan:Lee}. In the SZ case, we depend only on the well-characterised large-scale structure of the magnetic field within Coma~\cite{Bonafede:2010} and have a variety of high-precision data on the SZ effect within galaxy clusters (particularly Coma) available~\cite{dePetris:2002,Planck}. When studying realistic astrophysical environments, magnetic fields are usually present and are needed to account for energy losses occurring due to synchrotron radiation.

In this work, we use the SZ effect in a novel way to place constraints on PBH abundance. We shall consider the induced SZ effect from electrons created by PBHs via Hawking radiation, in galaxy clusters. In particular, we are looking for cases where the SZ effect from PBHs is brighter than the X-ray inferred electron population in Coma.

\section{Methodology}

Firstly, we determine the number of electrons a PBH produces via Hawking radiation. This radiation has a quasi-thermal spectrum~\cite{Corda:2012} with temperature $T$ and the standard thermal shape

\begin{equation}
\frac{d N}{d E} = \left(\frac{E}{8\pi k_B T}\right)^2 \frac{p_a}{2\pi} \left(\mathrm{e}^{\frac{E}{k_B T}}+1\right)^{-1}  \; ,
\end{equation}
where the numerical coefficients $p_a$ vary slightly for lower and higher-energy electrons
\begin{equation}
p_a = \left\{ \begin{array}{ll} 
27/h & E \geq 8 \pi k_B T \\
16/h & E < 8 \pi k_B T  \\ \end{array} \right. \; .
\end{equation}
Here, $dN/dE$ has units of inverse energy, inverse time. The fact that the temperature of the radiation is proportional to the average energy $\langle E\rangle$ of the electron means that an average galaxy cluster (with $k_B T = 10$ keV) can host a large number of energetic electrons, depending on the number of PBHs which undergo Hawking radiation in the beginning. To find the corresponding electron equilibrium distribution we need to solve the steady-state diffusion-loss equation
\begin{equation}
\vec{\nabla}\cdot\left(D(E,\vec{x})\vec{\nabla}\frac{d n_{e}}{d E}\right) + \frac{\partial}{\partial E}\left[b(E,\vec{x})\frac{d n_{e}}{d E}\right] + Q_{e}(E,\vec{x}) = 0 \; ,
\end{equation}
where $D$ is the diffusion function, $b$ is the loss function, and $Q_e$ is the ``source term" (i.e. the rate of electron injection by PBHs). The loss function is given by
\begin{equation}
\begin{aligned}
b(E) & = b_{\mathrm{IC}} E_{\mathrm{GeV}}^2 + b_{\mathrm{sync}} E_{\mathrm{GeV}}^2 B_{\mu\mathrm{G}}^2 \;\\ & + b_{\mathrm{Coul}} \overline{n}_{\mathrm{cm}3} \left(1 + \frac{1}{75}\log\left[\frac{\gamma}{\overline{n}_{\mathrm{cm}3}}\right]\right) \\& + b_{\mathrm{brem}} \overline{n}_{\mathrm{cm}3} E_{\mathrm{GeV}}\;,
\end{aligned}
\end{equation}
where $\gamma = \frac{E}{m_e c^2}$ with $m_e$ being the electron mass, $\overline{n}$ is the average gas density, and $\overline{n}_{\mathrm{cm}3} = \left(\frac{\overline{n}}{1 \; \mbox{cm}^{-3}}\right)$. The coefficients $b_{\mathrm{IC}}$, $b_{\mathrm{sync}}$, $b_{\mathrm{Coul}}$, $b_{\mathrm{brem}}$ are the energy-loss rates from ICS, synchrotron emission, Coulomb scattering, and bremsstrahlung. These coefficients are given by $0.25\times 10^{-16}(1+z)^4$ (for CMB target photons), $0.0254\times 10^{-16}$, $6.13\times 10^{-16}$, $4.7\times 10^{-16}$ in units of GeV s$^{-1}$. 

We specify $Q_e$:
\begin{equation}
Q_e = \frac{d N}{d E} \frac{\rho_{\mathrm{PBH}}}{M_{\mathrm{PBH}}} \; ,
\end{equation}
where the density profile is the DM distribution of the target object (i.e. the Coma cluster). If PBHs are to make up the entirety of DM, then they should be distributed in the same manner as the DM. A general profile is the Zhao-Hernquist case
\begin{equation}
\rho_{\mathrm{PBH}} (r) = \rho_s \left(\frac{r}{r_s}\right)^{-\alpha} \left(1 + \frac{r}{r_s}\right)^{-3 + \alpha} \; ,
\end{equation}
where $r$ is the radial coordinate and $r_s$, $\rho_s$, and $\alpha$ are the characteristic values that describe a given halo. For galaxy clusters, the diffusion length is much smaller than the scale of the cluster. Therefore, for the simplest case, we are able to ignore diffusion. This leaves only the loss-function $b(E)$ to be calculated, see~\cite{Beck:2019} for details and the definition of the loss-function (we use the properties of the Coma cluster from \cite{Bonafede:2010,chen:clusters2007}). For this case, our solution for the differential electron equilibrium distribution looks like
\begin{equation}
\frac{d n_{e}}{d E} (r,E) = \frac{1}{b(E)}  \int d E^{\prime} \, Q_e (r,E^{\prime}) \; .
\end{equation}
Integrating over the energy $E$ we find the electron distribution created by PBHs via Hawking radiation, where diffusion is insignificant
\begin{equation}
n_{e,\mathrm{PBH}}(r) = \int d E \frac{\rho_{\mathrm{PBH}}}{M_{\mathrm{PBH}} b(E)} \int d E^\prime \frac{\Gamma (E^\prime)}{\exp\left[E^\prime/T_{\mathrm{PBH}}\right]+1} \; .    
\end{equation}
The electron absorption probability is given by $\Gamma (E)$. We note that $\rho_{\mathrm{PBH}} = f \rho_{\mathrm{DM}}$, where $f$ is the amount of DM in the form of PBHs. The electrons, created by Hawking radiation, scatter CMB photons and lose energy. This will result in a change of energy and therefore a change in temperature. The total change in temperature requires we use
\begin{equation}
\Delta T (x) = y g_{sz}(x) \; ,
\end{equation}
where $x = \frac{h \nu}{k_B T_{\mathrm{CMB}}}$, $\nu$ is the frequency of interest and $g_{sz}$ is the spectral distortion function~\cite{Beck:2019}. The Compton-$y$ parameter is given by 
\begin{equation}
y = \sigma_T \int dl \, n_e (r) \frac{k_B T}{m_e c^2} \; ,
\end{equation}
where $l$ is the line of sight through the target halo. In the simplest case (the diameter line) $r = l$. Note, we will need to integrate our equilibrium solution $\frac{d n_{e}}{d E}$ over $E$ as well as $l$.

Next, we calculate the change in temperature due to a PBH population that constitutes all of the DM within the Coma cluster $\Delta T_{\mathrm{PBH}}$ and compare it to the data presented in \cite{dePetris:2002}. We may then use
\begin{equation}
f \Delta T_{\mathrm{PBH}} < \Delta T_{\mathrm{Coma}} \; ,
\end{equation}
to limit the allowed values of $f$ at 95\% confidence interval by using the $2\sigma$ upper limit on $\Delta T_{\mathrm{Coma}}$. 

Note that, in the case of Planck data~\cite{Planck}, we compare the SZ $y$ values in the centre of the cluster, as their temperature profile is not easily resolved by frequency (due to details of their analysis). 

\section{Results and discussion}

\begin{figure}[h]
\begin{minipage}{15pc}
\includegraphics[width=18pc]{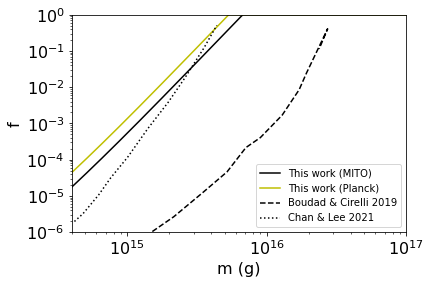}
\caption{\label{Fig1}Constraints on PBH abundance for the monochromatic mass function.}
\end{minipage}\hspace{3pc}%
\begin{minipage}{15pc}
\includegraphics[width=18pc]{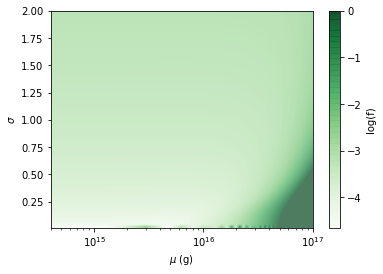}
\caption{\label{Fig2}Constraints on PBH abundance for the extended mass function.}
\end{minipage} 
\end{figure}

First, we will discuss the monochromatic case, where all the PBHs were created with the same mass. The most stringent constraints, in the studied mass range, for the monochromatic case are those by Boudad \& Cirelli 2019, see Figure \ref{Fig1}. The constraints of Chan \& Lee 2020 and those in this work are a lot weaker, however, we found that our results are comparable to Chan \& Lee 2020. In Figure \ref{Fig1}, our constraints are given by the solid black and yellow lines. The yellow line is associated with the Planck data and is thus likely more representative of an accurate physical picture due to the superiority of the instrument. The black line corresponds to the MITO data \cite{dePetris:2002}, which resulted in slightly better results than Planck. For both our data sets we see that, while our results are slightly weaker for masses below $10^{15}$ g, we can constrain PBHs of larger mass better than Chan \& Lee 2020. Importantly, we do not suffer from their systematic uncertainties in the magnetic field modeling and our limits are 95\% confidence interval, whereas theirs are naive upper limits. 

Additionally, from Figure \ref{Fig1}, we can see that our constraints are not as stringent as those from Boudad \& Cirelli 2019. Low-energy electrons are difficult to study since they are affected by the sun as they move through space. Recently, however, the Voyager $1$ spacecraft left the sun's heliopause thereby lifting the limitation on studying low-energy electrons. Boudad \& Cirelli 2019 were then able to find constraints that were based on a more comprehensive study. 

Next, we consider the extended mass distribution. There are three commonly used: log-normal, power-law, and critical collapse. In this work, a log-normal mass distribution is used. It is given by
\begin{equation}
\frac{dn}{dM} = \frac{f}{\sqrt{2\pi}\sigma M}\exp\left({-\frac{\log^2(M/\mu)}{2\sigma^2}}\right) \; ,
\end{equation}
where each set $(\mu,\sigma)$ provides a model represented by part of a parameter space corresponding to an $f$. Here, $\mu$ is the mass for which the density is a maximum and $\sigma$ is the width around the peak. For a large $\mu$ and small $\sigma$, for example, we can see that the parameter space is largely a dark green. Dark green indicates a large fraction $f$, i.e. close to $f\sim 1$. Considering the entire parameter space, we notice that dark green covers a very small patch. Therefore, the darker the green, the more poorly constrained the model. This is encouraging since most of the models (i.e. the parameter space) are well-constrained as can be seen by lighter greens. For the extended case $3\times 10^{15}$g corresponds to the range $f\in \left[10^{-4},10^{-3}\right]$ g across all values of $\sigma$. This range is much smaller than unity, so it reasonably constrains the PBH abundance. Interestingly, our results for extended distributions are slightly more comparable to those presented by Boudad \& Cirelli 2019 than in the monochromatic case. 

Boudad \& Cirelli 2019 used the full, general diffusion-convection-reacceleration model to study propagation \cite{Boudaud:Cirelli}. This yielded well-constrained PBH-DM abundances due to Voyager 1's access to low-energy electrons. On the other hand, Chan \& Lee 2020 considered synchrotron emission from the galactic centre, where the magnetic field is quite uncertain~\cite{Bringmann:2014} and is highly important to their results. In our case, the results are weaker than Boudad \& Cirelli 2019 but provide a novel and robust means to derive constraints on PBH populations, without the serious systematic concerns of the method of Chan \& Lee 2020. One caveat on these results is that we have used the non-relativistic SZ effect throughout. Whereas, the high temperatures for lower mass PBHs would necessitate the considerably more complex relativistic approach. This will be explored in future work and may enhance the limits for PBH masses below $10^{15}$g.

\section{Conclusion}

We present a novel means of constraining the abundance of PBHs as DM by considering the electrons produced via Hawking radiation of the PBHs. These energetic electrons provided the necessary electron population required to scatter photons from the CMB. We used the SZ effect to calculate a temperature change due to a population of PBHs and compared this to SZ data for the Coma cluster. Constraints on $f$, which is the amount of DM consisting of PBHs, could be placed with a 95\% confidence level. We considered two mass functions. The monochromatic results were comparable to Chan \& Lee with a slight advantage toward higher masses. The extended case placed better constraints. The resulting parameter space was mostly light green giving $f<10^{-2}$ and only a small corner with $f\sim 1$ was poorly constrained. One advantage of this method is that it is more robust than the synchrotron approach and should be applicable in a larger number of environments.

\ack
JT acknowledges the support of the SARAO post-doctoral fellowship initiative. 

\section*{References}

\medskip

\bibliographystyle{iopart-num}
\bibliography{bib_SAIP2022}

\end{document}